# Improved quantum sensing with a single solid-state spin via spin-to-charge conversion

J.-C. Jaskula[1,2], B. J. Shields[2,*], E. Bauch[2], M. D. Lukin[2], A. S. Trifonov[1,2,3] and R. L. Walsworth[1,2,4]

[1] Harvard-Smithsonian Center for Astrophysics, Cambridge, Massachusetts 02138, USA

[2] Department of Physics, Harvard University, Cambridge, Massachusetts 02138, USA

[3] Ioffe Physical-Technical Institute RAS, Saint Petersburg, Russia

[4] Center for Brain Science, Harvard University, Cambridge, Massachusetts 02138, USA

**Abstract**

Efficient optical readout of a single, solid-state electronic spin at room temperature is a key challenge for nanoscale quantum sensing. Here we apply the technique of spin-to-charge conversion to enhance the optical spin-state readout of a single Nitrogen-Vacancy (NV) color center in room temperature diamond, with no degradation in the NV spin coherence time. We demonstrate an order-of-magnitude improvement in spin readout noise per shot and about a factor of five improvement in AC magnetometry sensitivity, compared to the conventional NV spin-state optical readout method. This improvement is realized in a widely-applicable bulk diamond system. We show that selecting for successful charge state initialization leads to possible further improvement in sensitivity. This technique is well suited to sensing applications involving low duty cycle pulsed signals, e.g., in biomagnetometry, where long deadtimes demand optimized sensitivity per shot.

## 1. Introduction

Quantum defects in solids are emerging as the sensors of choice for detecting micrometre to nanometre phenomena in a wide range of systems in both the physical and life sciences. The Nitrogen-Vacancy (NV) color center in diamond is a prime example of such a solid-state quantum defect, combining long-lived electronic spin coherence with the ability to prepare and read out the spin state optically [1]. For example, NV centers have been used for high-spatial resolution sensing of magnetic fields in cell biology [2], bioassays [3], neuroscience [4], nanoscale NMR [5-7], and condensed matter physics [8-11]. An important figure of merit for this system is the spin readout noise per shot, specifying the accuracy with which the final state of the electron spin evolution can be measured in a single measurement attempt (shot). Current room temperature NV experiments rely on a spin-dependent fluorescence signal that is restricted to a short detection window (~250 ns), after which the spin is optically pumped into the $m_S = 0$ state, such that the spin-state-dependent fluorescence contrast is typically limited to ~25% for a single NV center [1]. Recently, a new NV spin readout technique based on spin-to-charge conversion (SCC) that overcomes this limitation for NV centers was demonstrated in diamond nanobeams [12], in which the single NV fluorescence is enhanced by an order of magnitude compared to NV centers in bulk diamond. A similar protocol has also been demonstrated using high-intensity, near-infrared laser pulses and single NV centers under solid immersion lenses to achieve a spin-to-charge mapping via selective ionization of the singlet manifold [13]. Here, we show that SCC readout provides an order of magnitude improvement in single NV spin readout noise per shot in bulk diamond, without any influence of the charge state preparation on the NV spin coherence time. We then use SCC readout to demonstrate an improvement in single NV AC magnetometry sensitivity of about a factor of five, which is also about five times above the spin projection noise limit, similar to previous results for a single NV in nanobeams [12].

---

[*] Department of Physics, Basel University, Basel, Switzerland.

The conventional mechanism for room temperature optical readout of the NV electron spin (spin-1) is based on a spin-dependent intersystem crossing from the optical excited state into a manifold of optically dark singlet states. Under excitation with 532nm light, an NV center initially in $m_S = 1$ is shelved into the metastable singlet level with about a 50% probability [14] and remains dark under subsequent excitation during the singlet state lifetime; whereas an NV center initially in $m_S = 0$ will continue to cycle and scatter photons during this readout time. These spin-state-dependent pathways induce repolarization of an NV center to the $m_S = 0$ ground state and constrains the detection window duration to ~250 ns at room temperature. Consequently, the fluorescence collected from a single NV center in bulk diamond is limited to about 0.022 photons per readout with a low contrast (Figure 1(a)). Thus, effective readout of the NV spin state can only be achieved statistically after many averages. Standard approaches to overcome this limitation involve using macroscopic ensembles of NVs [15-17] or photonically enhanced diamond nanostructures [21-23], in order to boost the fluorescence signal from the NV centers. Such techniques have their limitations, namely reduced spatial resolution and coherence time in the case of ensemble measurements, or limited sensing area and scalability in the case of photonic nanostructures. Therefore, techniques for improved readout of single NV centers in bulk diamond are of great interest.

## 2. Spin-to-charge state conversion mechanism

A complementary approach is to develop readout techniques that transfer the spin state to more robust degrees of freedom than the relatively short-lived singlet. One such technique involves the transfer of spin information to the NV nuclear spin, which can subsequently be repetitively read out multiple times through the NV electronic spin [18]. This technique is capable of approaching the spin-projection noise limit, but has the disadvantage of requiring strong magnetic fields to split the nuclear spin levels, limiting its applicability. The present work deploys a spin-to-charge conversion (SCC) mechanism [12] that works by transferring the spin state to the charge state of the NV, followed by a high-fidelity charge state measurement. This method has the advantage of being able to reach spin readout noise levels as low as twice the spin projection noise level, while being fully optical and easily incorporated in a confocal microscope and potentially with ensembles of NV centers.

NV centers are found mainly in two charge configurations [19], $NV^0$ and $NV^-$ (figure 1(b)), which can be switched via a photo-ionization process [20]. Moreover, the NV center charge state can be read out in a single shot with a laser beam at 594nm, which efficiently excites $NV^-$ but only weakly $NV^0$, producing a high fluorescence contrast between the two charge states [21-23] (Figure 1(c)). For high readout fidelity, photo-ionization must be suppressed during readout, requiring low laser powers (as low as a few microwatts) and long readout times (from tens of microseconds to few milliseconds, depending on the desired single shot charge state readout fidelity). To switch between the charge states, a green laser beam (532 nm) initializes the charge state of the NV center preferentially to $NV^-$ with 70% probability; while a 10-nanosecond pulse of red light (~637 nm) almost ideally ionizes the NV center from $NV^-$ to $NV^0$ via a two-step process that occurs between the NV center ground state and the conduction band.

The ionization process under 637nm illumination can be made spin-state-dependent by first shelving one spin state into the metastable singlet level, for which ionization is supressed. In particular, the $m_S = 1$ population can be protected from ionization by first transferring it with a 594nm, 50ns-long pulse into the singlet state before ionizing the remaining triplet state population (mainly $m_S = 0$) with an intense 637nm pulse. The 637nm light does not excite the singlet state, so the ionization process is blocked for $m_S = 1$, and the NV remains in the negative charge state. We implement this spin-to-charge conversion with the sequence depicted in figure 1(d). We plot the photon number distributions in figure 1(e) for the two cases where the NV center is prepared in $m_S = 0$ (no microwave pulse) and $m_S = 1$ (microwave π pulse), illustrating our SCC efficiency. As expected, $m_S = 0$ is converted almost entirely to $NV^0$, while $m_S = 1$ remains in NV⁻ to a large degree. From the mean values of the two distributions ($<N_{ph}>_{m=0}$ = 2.4 and ($<N_{ph}>_{m=1}$ = 3.8), we extract a spin-state contrast of 36%, which is larger than for the conventional (singlet-state) readout scheme (by about a factor of 1.5x) due to the fact that the ionization occurs when the population of the singlet state is maximum. More importantly, the robustness of the charge states allows for the collection of an arbitrary number of photons in a single shot (typically, about 5 photons in a 1 ms time window). This increased photon number largely eliminates the contribution of photon shot noise in the measurement, limiting the SCC method to spin-projection noise and conversion noise during the spin-to-charge mapping.

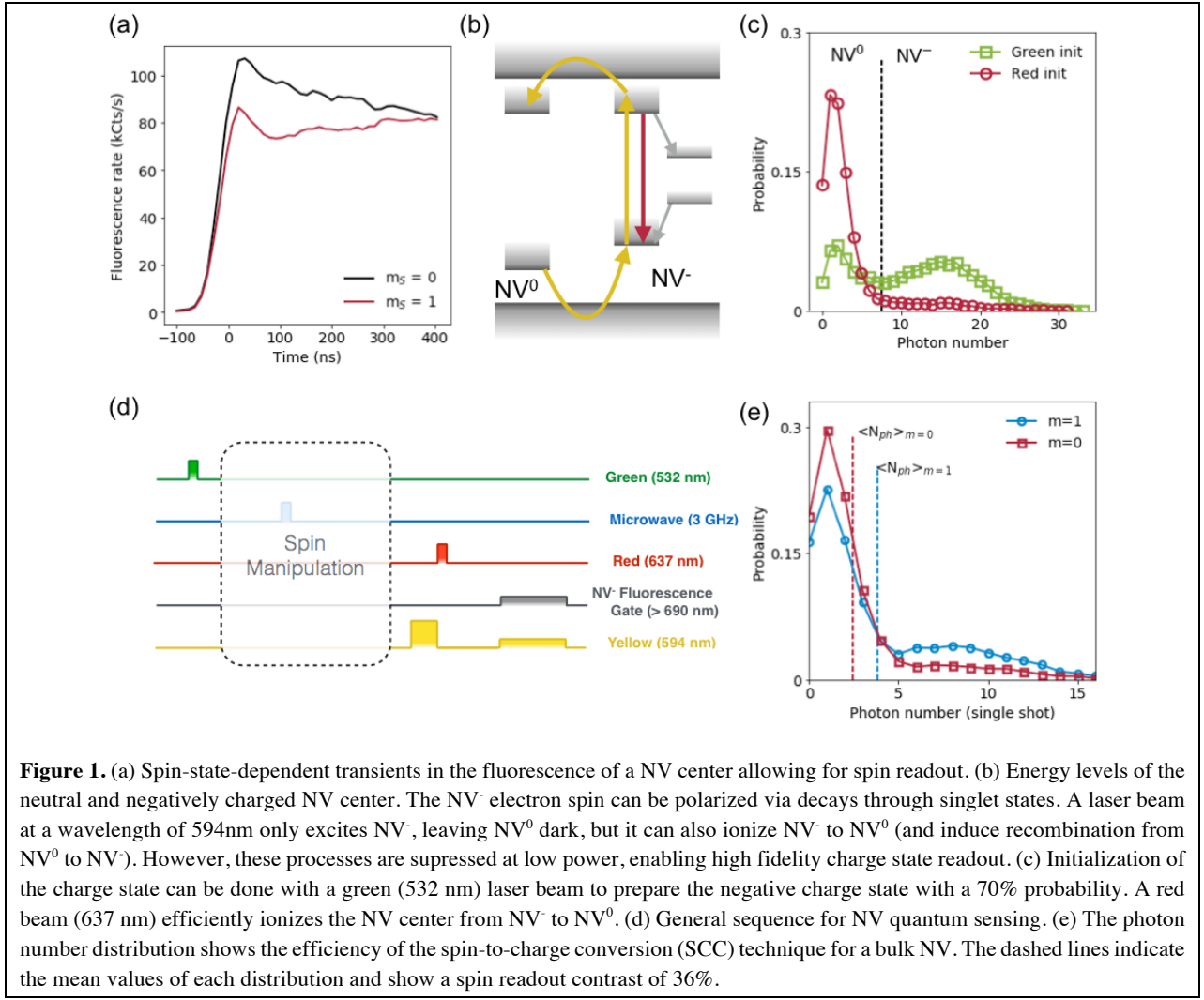

**Figure 1.** (a) Spin-state-dependent transients in the fluorescence of a NV center allowing for spin readout. (b) Energy levels of the neutral and negatively charged NV center. The NV⁻ electron spin can be polarized via decays through singlet states. A laser beam at a wavelength of 594nm only excites NV⁻, leaving NV⁰ dark, but it can also ionize NV⁻ to NV⁰ (and induce recombination from NV⁰ to NV⁻). However, these processes are supressed at low power, enabling high fidelity charge state readout. (c) Initialization of the charge state can be done with a green (532 nm) laser beam to prepare the negative charge state with a 70% probability. A red beam (637 nm) efficiently ionizes the NV center from NV⁻ to NV⁰. (d) General sequence for NV quantum sensing. (e) The photon number distribution shows the efficiency of the spin-to-charge conversion (SCC) technique for a bulk NV. The dashed lines indicate the mean values of each distribution and show a spin readout contrast of 36%.

We define the spin readout noise per shot $\sigma_R$ as the ratio between the measured spin noise and the spin projection noise $\sigma_{SPN} = \sqrt{p_0(1-p_0)}$, where $p_0$ is the probability of detecting one projection, e.g., $m_S = 0$, of our effective two-level spin system. The spin readout noise can be estimated directly from the collected photon distribution for each class of spin. In the case of SCC readout, a perfect mapping would link the spin state $m_S = 0$ and $m_S = 1$ to the neutral charge state NV⁰ and negatively charged state NV⁻, respectively. We define errors $\epsilon_0$ and $\epsilon_1$ to be the probabilities that the detected charge state deviates from such a perfect mapping. Then we can recast the spin readout noise per shot expressed in [12] as $\sigma_R = \frac{\sqrt{1-(\epsilon_0-\epsilon_1)^2}}{1-(\epsilon_0+\epsilon_1)}$. These errors can be straightforwardly estimated from figure 1(e) by choosing a threshold (here equal to 5 photons) that distinguishes the charge states from each other as depicted in figure 1(c). In the experiments reported here, we typically obtain $\epsilon_0 \sim 0.14$ and $\epsilon_1 \sim 0.69$, which results in $\sigma_R \sim 5$. For comparison, conventional NV spin-state readout is limited by photon shot noise [1, 12]. The spin readout noise per shot consequently becomes $\sigma_R \approx 2(V\sqrt{n})^{-1} \approx 54$, where $n \approx 0.022$ is the average mean photon number per detection window and $V \approx 25\%$ is the contrast for single NV confocal detection (Figure 1(a)).

## 3. Experimental Setup

The experimental setup is a confocal microscope adapted for NV spin-state measurements via both the SCC and conventional fluorescence techniques. Three laser beams (at 532, 594, and 637 nm) are coupled into single mode optical fibers to improve their spatial mode, and are then combined together with dichroic mirrors. A DPSS laser is used together with an acousto-optic modulator (AOM) to generate the green (532 nm) laser pulse. The protective yellow (594 nm) laser pulse and charge readout are both made with the same laser (HeNe 1.5mW) and AOM. To switch quickly from a relatively high yellow laser power (500 $\mu$W) for the protective step to a low power (5 $\mu$W) for the charge-state readout, two RF control signals are generated with voltage-controlled oscillators and controlled with RF switches. The red (637 nm) ionizing optical pulses are generated with a diode laser (HL63133DG, Thorlabs, 170 mW CW at 637nm). The optical pulses are controlled with a pulse generator (AVO-2L-C-GL2, Avtech, 500 ps rise time, 2A), resulting in square pulse widths between 4 and 30 ns. The three laser beams are focused on a single NV in bulk diamond with a 1.45 NA, oil-immersion objective. The NV fluorescence (~90 kcounts/s) is collected through the same objective and then detected with an avalanche photodiode. A 150 $\mu$m pinhole restricts the longitudinal point-spread function to 500 nm and avoids out-of-focus background fluorescence from other NV centers.

## 4. Application to magnetic sensing

*4.1 AC Magnetometry*. As a proof of concept of the performance of the SCC method for quantum sensing, we use the single NV center as a nanoscale AC magnetic field sensor. Oscillating magnetic fields can be probed by applying a synchronized Hahn-echo sequence ($\pi/2$ - $\tau/2$- $\pi$ - $\tau/2$ - $\pi/2$). During the period $\tau$, the NV center's quantum state evolves under the influence of the AC magnetic field as well as environmental noise. This evolution is manifested by a phase accumulation that is converted into an optically-readable spin state population difference. Because of the noise environment, such phase accumulation is limited to the characteristic coherence time $T_2$. For a single NV, $T_2$ can approach one millisecond [24, 25] via the implementation of dynamical decoupling sequences that act as filter functions to suppress the effects of noise fluctuations. We drive the NV center's electronic spin into a superposition of $m_S = 0$ and $m_S = 1$ that acquires a relative phase scaling linearly with the magnetic field amplitude $\phi = \frac{2\mu_B}{\pi\hbar} B \tau = \alpha B \tau$ [26, 27]. This measurement protocol results in a rotation of the Bloch vector in the equatorial plane, which is observed as a sinusoidal oscillation of the NV fluorescence signal with AC magnetic field amplitude, as the phase is transferred to a population difference (figure 2(a)).

Figure 2(b) is a zoom around the point of zero AC magnetic field amplitude, where the NV magnetometer sensitivity $\eta(\tau)$ is maximum, which we estimate by [1]:

$$\eta(\tau) = \sigma_R \frac{e^{\left(\frac{\tau}{T_2}\right)^p}}{\alpha \sqrt{\tau}} \sqrt{\frac{t_{init}+\tau+t_{ro}}{\tau}}. \qquad (1)$$

Here, $\sigma_R$ is the spin readout noise per shot, normalized to the spin projection noise, $\alpha$ is the scaling factor between the phase acquired and the field measured, and p is an exponent containing information related to the spin bath [28]. $\sqrt{(t_{init} + \tau + t_{ro})/\tau}$ is a time penalty, where the initialization time and the readout time are denoted respectively $t_{init}$ and $t_{ro}$. For our demonstration experiment, we find an improvement of a factor 5 in AC magnetic field sensitivity of the SCC method, 9(1) nT/$\sqrt{\text{Hz}}$, over the conventional readout technique, 45(12) nT/$\sqrt{\text{Hz}}$.

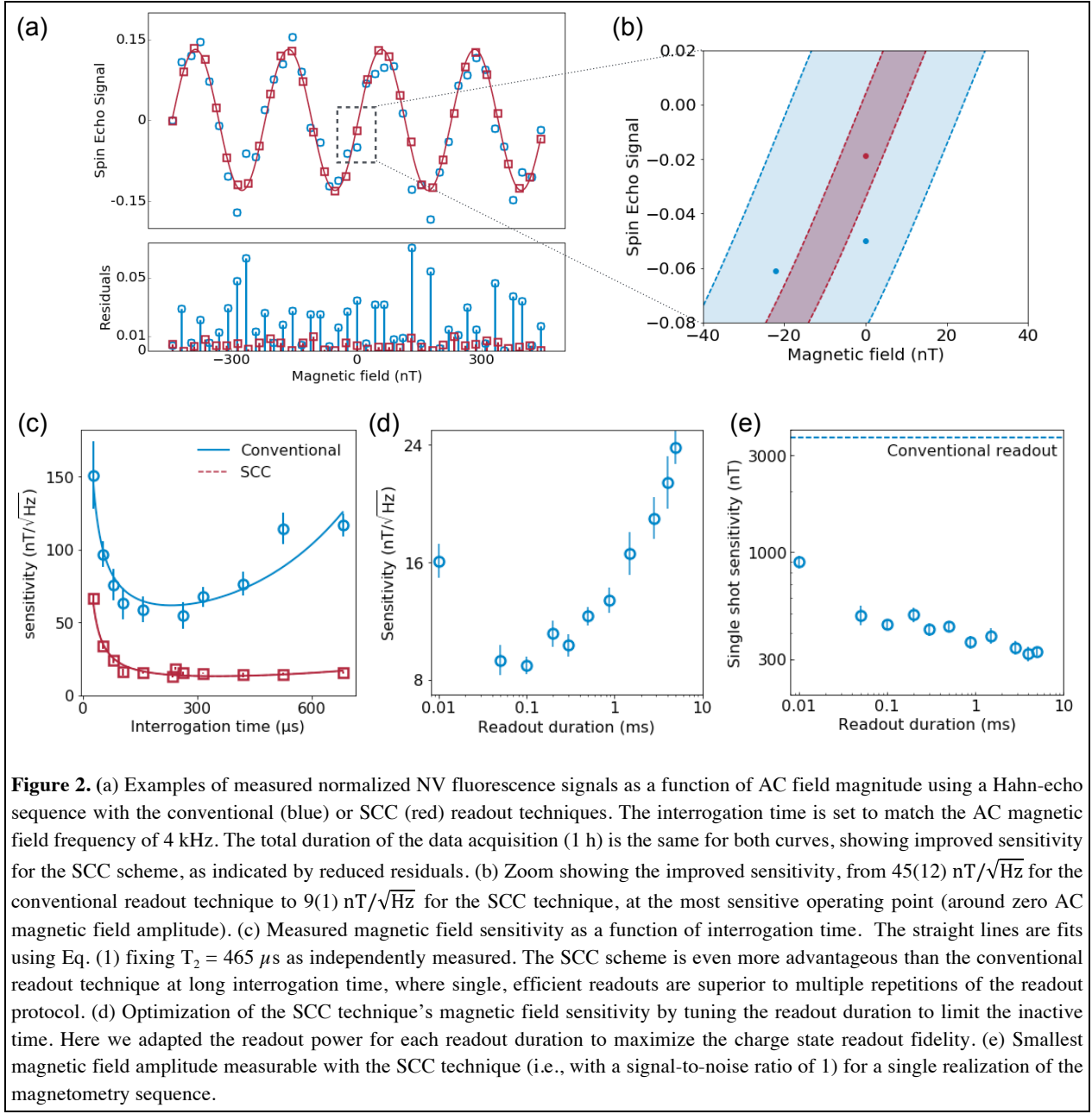

**Figure 2.** (a) Examples of measured normalized NV fluorescence signals as a function of AC field magnitude using a Hahn-echo sequence with the conventional (blue) or SCC (red) readout techniques. The interrogation time is set to match the AC magnetic field frequency of 4 kHz. The total duration of the data acquisition (1 h) is the same for both curves, showing improved sensitivity for the SCC scheme, as indicated by reduced residuals. (b) Zoom showing the improved sensitivity, from 45(12) nT/$\sqrt{\text{Hz}}$ for the conventional readout technique to 9(1) nT/$\sqrt{\text{Hz}}$ for the SCC technique, at the most sensitive operating point (around zero AC magnetic field amplitude). (c) Measured magnetic field sensitivity as a function of interrogation time. The straight lines are fits using Eq. (1) fixing $T_2 = 465\ \mu$s as independently measured. The SCC scheme is even more advantageous than the conventional readout technique at long interrogation time, where single, efficient readouts are superior to multiple repetitions of the readout protocol. (d) Optimization of the SCC technique's magnetic field sensitivity by tuning the readout duration to limit the inactive time. Here we adapted the readout power for each readout duration to maximize the charge state readout fidelity. (e) Smallest magnetic field amplitude measurable with the SCC technique (i.e., with a signal-to-noise ratio of 1) for a single realization of the magnetometry sequence.

We next investigate the sensitivity improvement provided by the SCC scheme for different evolution times $\tau$. Increasing this time results in a stronger integrated readout signal, but is only favorable up to the point where spin decoherence reduces the measurement contrast. According to equation (1), there is an optimum of sensitivity at $\tau = T_2/2 = 232\ \mu$s for the conventional readout scheme (for which the initialization and readout time are negligible and $p \approx 1$), as is clearly visible in figure 2(c). The SCC technique performs noticeably better than conventional readout over a wide range of evolution times between 50 $\mu$s to 700 $\mu$s. As $\tau$ increases and becomes significantly longer than the readout time, the gain must approach its maximum value, $\sigma_R^{\text{SCC}}/\sigma_R^{\text{stand.}}$. At the other extreme, we expect that short spin readout times will prevail when $\tau$ is short since they allow for fast repetition of quantum sensing sequences. We find that this regime is not reached yet at evolution times as short as 50 $\mu$s, where the sensitivity improvement is still in favor of the SCC technique by a factor of two (figure 2(c)). From the parameters of each fit, we determine that the spin readout noise per shot $\sigma_R$ for conventional readout is $\sigma_R^{conven.} \approx 60$ and $\sigma_R^{SCC} \approx 5$ for SCC readout, confirming the values reported above. Interestingly, the fitted p parameter in (1) is different for the two readouts ($p_{SCC} = 1.01$ and $p_{conven} = 1.33$), so that the sensitivity improvement is not yet saturated for 700 $\mu$s evolution time. Moreover, we note that AC magnetic field sensitivity at high frequency can be improved by dynamical decoupling sequences, which both move the filter function's central frequency to higher values (allowing long accumulation times even for fast oscillating fields), and extend the coherence time.

Since the time penalty term in (1) becomes significant only for a readout time comparable to or longer than the interrogation time, the SCC sensitivity improvement can be optimized by increasing the readout time (and consequently decreasing $\sigma_R$) while keeping the time penalty negligible. We demonstrate this behaviour explicitly in figure 2(d) by measuring the sensitivity for different readout durations. At the evolution time where the sensitivity is optimal for conventional readout, we measure an SCC sensitivity of 9(1) nT/$\sqrt{Hz}$ for $t_{RO} = 40$ $\mu$s. On the other hand, the duty cycle of quantum sensors can also be limited by an external parameter, e.g., the repetition rate of the recorded events. In such scenarios, the correct figure of merit is the single shot magnetic field sensitivity. In particular, one can take full benefit of long readout time windows of a few milliseconds to improve the performance of the spin readout and consequently of NV magnetometers using the SCC technique. For example, as shown in Figure 2(e), we realize a magnetic field amplitude uncertainty of 307(29) nT in a single SCC measurement with a 5 ms readout time, for which the photon shot noise is nearly totally suppressed. This uncertainty is an order of magnitude smaller than with conventional NV spin readout.

*4.2 Charge state initialization monitoring.* The 30 $\mu$s-long green initialization leaves the NV center in its neutral charge state one third of the time, allowing for a substantial improvement in the SCC technique performance by detecting aberrant initializations prior to the coherent microwave manipulation. This protocol can be performed with charge-state readout capability with either a 'try till success' polarization technique or by post-selecting on the initial charge state being NV⁻. Although the former offers the best sensitivity, it also requires fast readout electronics to compute the result of the first charge state readout. We implemented the latter by reading out the charge state of the NV center with a 10 ms long yellow readout pulse. The duration is chosen to limit ionization during this step, although some remains in practice, as one can see that the spin readout noise per shot increases to about 8. In figure 3(a), we plot the distribution of photons obtained from the first readout gate (blue curve). The post-selection is conditioned on having acquired a number of photons during the first charge state readout pulse that is strictly higher than a certain threshold (6 photons in figure 3(a), filled blue area). Because we mainly reject sequences where the NV center is in NV⁰, the distinguishability between the spin states is enhanced (Figure 3(b)). To further demonstrate the improvement due to post-selection, we measure the readout noise for various thresholds. We see in figure 3(c) that $\sigma_R$ decreases faster at low threshold because we remove measurements where the NV center could not be used to sense any magnetic fields. After a threshold that corresponds to the visual limit between NV⁰ and NV⁻, the slope becomes less steep as we discard good measurements. In parallel, we estimate the cost of rejecting measurements by computing an effective sequence time. The sensitivity can be improved if $\sigma_R$ decreases faster than the square root of the effective sequence time increases (eq. (1)). We show that this is indeed the case for thresholds smaller than 6 photons per readout pulse, with an improvement of about 5% (Figure 3(d)), limited mainly by the long duration of the first charge state readout.

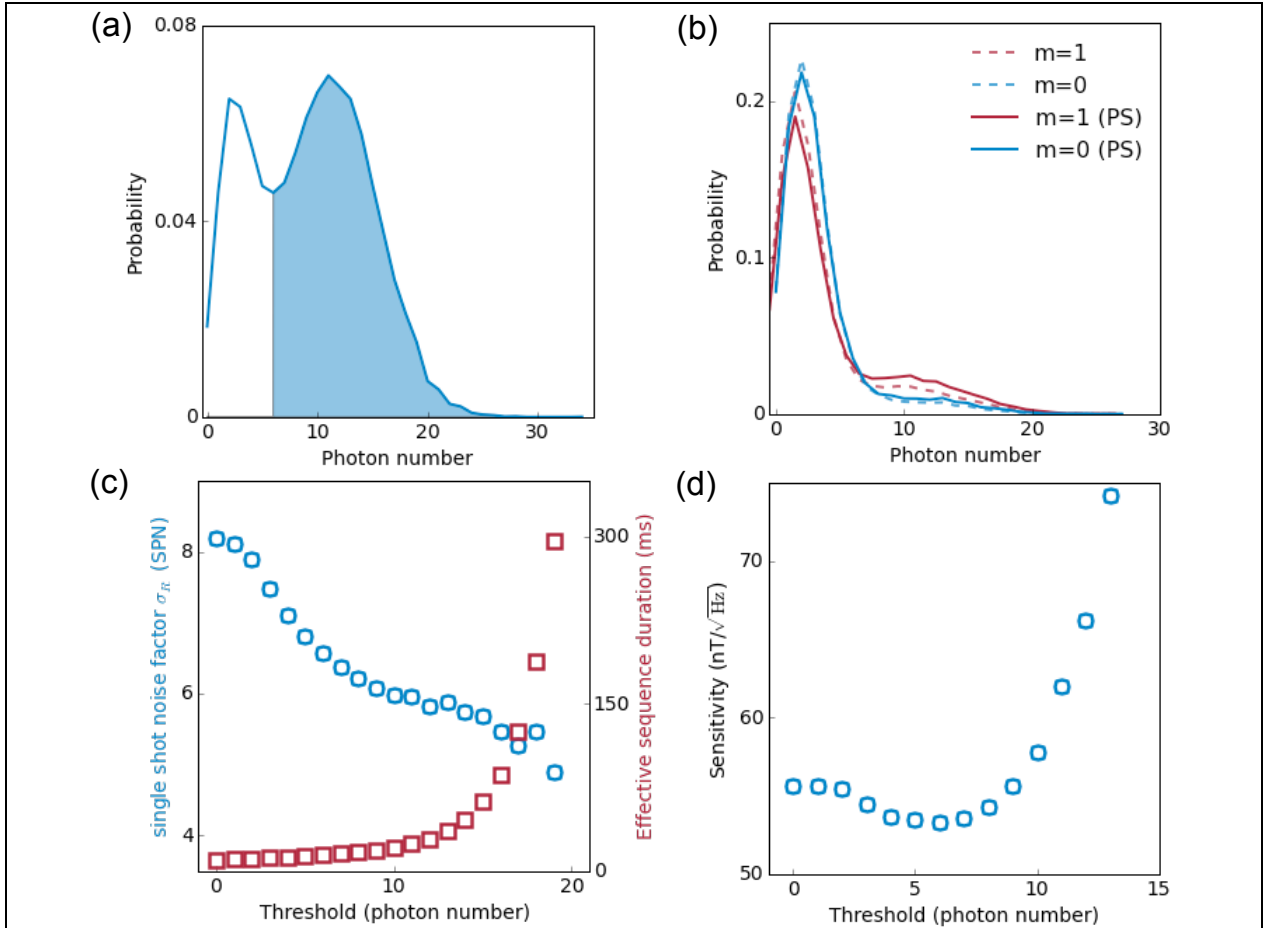

**Figure 3.** Post-selection enhancement of SCC technique. (a) Photon distribution recorded in an initial charge readout pulse prior to microwave manipulation. The distribution of all sequences (solid line) shows that a significant fraction of events is executed on the neutral charge state and can be discarded to keep mainly results obtained with NV⁻ (filled area). (b) Photon distributions after post-selecting for NV⁻ (solid lines) have greater distinguishability than the distributions for all events (dashed lines). The Kolmogorov–Smirnov test gives a distinguishability of 0.12 against 0.08 for no post-selection. (c) Spin readout noise per shot as a function of threshold, and effective sequence time. Eliminating initial NV⁰ events improves the efficiency of SCC mapping at the cost of increasing the sequence duration. (d) AC magnetic field sensitivity for different thresholds, showing a 5% improvement in sensitivity by post-selecting for the initial NV charge state. The length of both charge-state readout windows (20 ms in total) limits the absolute sensitivity to 53(16) nT/√Hz. Alternatively, very short time windows that would allow for the detection a single photon would be enough to determine the charge state of the NV center.

## 5. Conclusions

In this work, we show that the spin-to-charge conversion (SCC) technique, together with single-shot charge-state readout, provides about a factor of five improvement in AC magnetic field sensitivity for a single NV center in bulk diamond. This readout scheme is not only beneficial to AC magnetometry but also to any kind of quantum sensing measurement that benefits from long coherence time, such as $T_2$-limited thermometry [29] or ancilla-assisted DC magnetometry [30]. Furthermore, for typical NV spin dephasing times observed in isotopically purified $^{12}C$ samples with low Nitrogen concentration ($T_2^* \sim 50 \mu$s) [31], SCC readout should be applicable to Ramsey-like DC magnetometry with a two-fold sensitivity improvement. Moreover, SCC readout is well suited to detecting pulsed magnetic fields, for example those created by neuron action potentials with typical pulse timescales of milliseconds and delays between pulses of 10-100 milliseconds [4]. In such cases, it is not necessary to optimize the duty cycle of the sensor and one can take advantage of dead times to increase the fidelity of the readout. In particular, SCC sensitivity is expected to improve by an order of magnitude compared to conventional readout, down to ~300 nT for a single realization of the experimental protocol. Finally, because the neutral charge state is a dark state, the SCC readout scheme is also well suited for NV ensemble measurements and superresolution techniques [32, 33], which both suffer from background fluorescence. Indeed, off-axis NV centers are not affected by the microwave pulses and are consequently ionized during the spin to charge mapping.

**Acknowledgements**
This material is based upon work supported by, or in part by, the U. S. Army Research Laboratory and the U. S. Army Research Office under contract/grant number W911NF1510548, the DARPA Quantum Assisted Sensing And Readout (QuASAR) program (contract # HR0011-11-C-0073), the CUA, the Gordon and Betty Moore Foundation, the Vannevar Bush Fellowship, and the NSF EPMD, PoLS, and INSPIRE programs.

**Appendix A. Coherence time measurement**

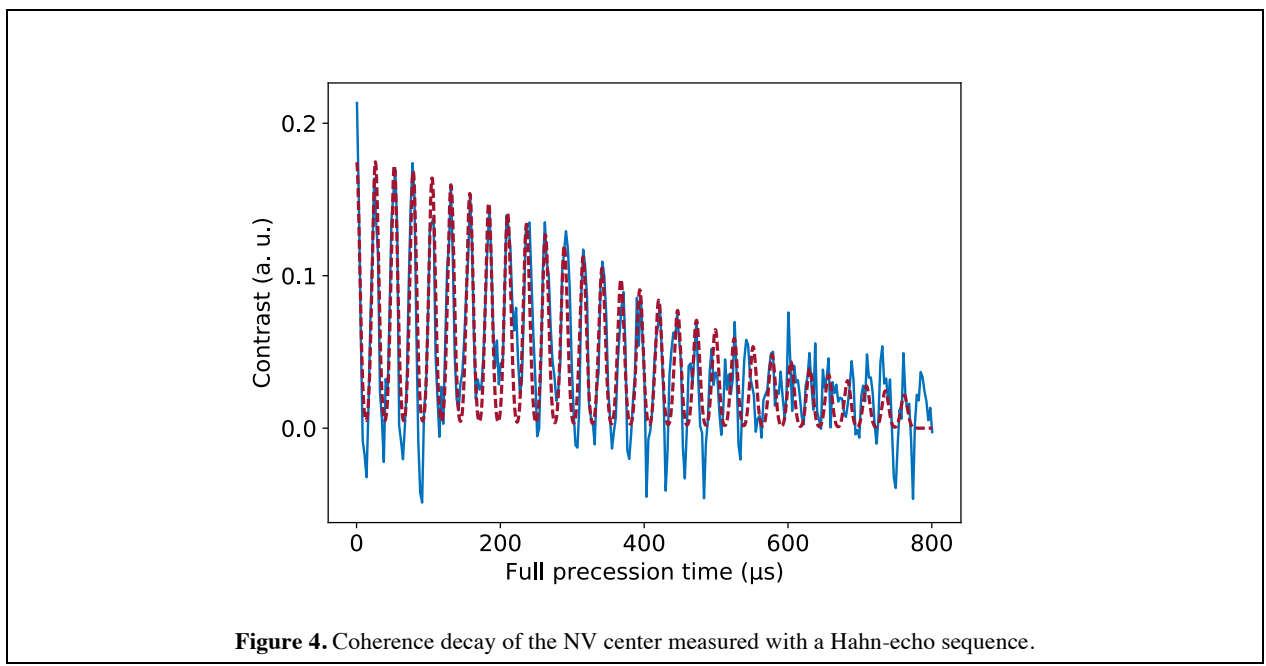

**Figure 4.** Coherence decay of the NV center measured with a Hahn-echo sequence.

In the absence of an applied AC magnetic field, the single NV coherence time $T_2$ can be measured using the same Hahn-echo sequence as described in the main text. As we vary the free evolution time $\tau$, we find that the superposition of $m_S = 0$ and $1$ decays to an incoherent state in a time $T_2 = 461.5$ $\mu$s. Due to the presence of $^{13}$C atoms in the diamond lattice (at the natural abundance of 1.1%), the coherence of the NV center is modulated by revivals that occur at multiples of the Larmor precession time $t_{rev} = 26.28$ $\mu$s. AC magnetometry can only be done at evolution times that corresponds to the different revivals. Consequently, the optimal experimental sensitivity is achieved for $\tau_{opt} = 223.34$ $\mu$s ~ $T_2/2$.

**Appendix B. Charge state readout of shallow NV centers**
Shallow implanted NV centers are a promising modality for nanoscale magnetic resonance imaging and single molecule detection due to the strong dipolar and hyperfine interactions with electronic and nuclear spin species located on the diamond surface [34-36]. Adversely, surface effects tend to shorten the coherence time of shallow NVs, typically to tens of microseconds, and could also potentially modify their charge state dynamics. We show in figure 5 that we see no change in the charge state dynamics of shallow NV centers; and also that they can still be read out with high fidelity. We plot in figure 5(a) a typical time trace of the observed fluorescence, under constant 594nm light illumination, from a single NV center implanted 3nm below the surface (measured by a nanoscale NMR technique [37]) exhibiting the expected behaviour: i.e. ionization and recombination between charge state $NV^0$ (low fluorescence level) and $NV^-$ (high fluorescence level).

The histogram of such time traces is plotted in figure 5(b) and displays two peaks associated with these two levels of fluorescence. We perform a numerical fit based on the master equation that describes the behaviour of the charge state under a 594nm light illumination [12] and extract the fluorescence rates $\gamma_0 = 200$ Hz and $\gamma_- = 1.3$ kHz, the ionization rate $g_0 = 45$ Hz, and the recombination rate $g_- = 6$ Hz for an excitation power of 280 $\mu$W. These rates are very similar to those measured with NV centers in bulk diamond and indicate that the SCC readout will provide a similar improvement with shallow NV centers as that reported in the main text for deep NVs.

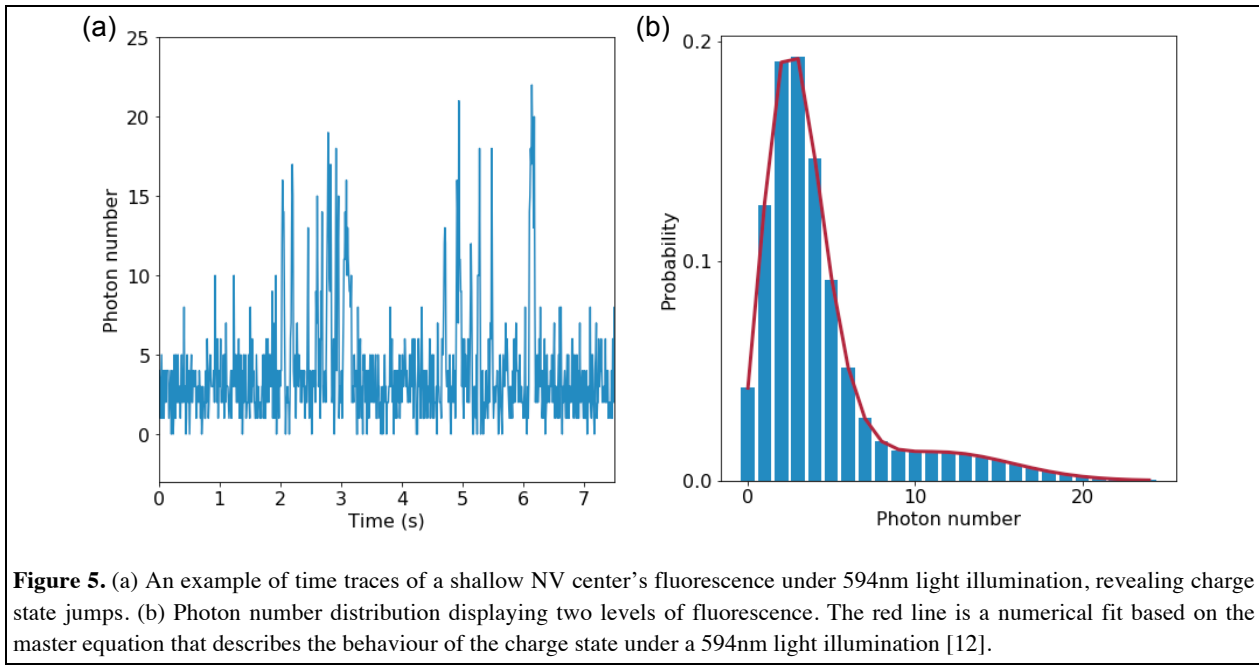

**Figure 5.** (a) An example of time traces of a shallow NV center's fluorescence under 594nm light illumination, revealing charge state jumps. (b) Photon number distribution displaying two levels of fluorescence. The red line is a numerical fit based on the master equation that describes the behaviour of the charge state under a 594nm light illumination [12].